# Highly repeatable nanoscale phase coexistence in vanadium dioxide films

T. J. Huffman[1], D. J. Lahneman[1], S. L. Wang[1], T. Slusar[2], Bong-Jun Kim[2], Hyun-Tak Kim[2,3] and M. M. Qazilbash[1*]

[1]Department of Physics, College of William and Mary, Williamsburg, VA 23187-8795, USA

[2]Metal-Insulator Transition Lab, Electronics & Telecommunications Research Institute, Daejeon 34129, Republic of Korea

[3]School of Advanced Device Technology, University of Science & Technology, Daejeon 34113, Republic of Korea

* Corresponding author: mumtaz@wm.edu

**The metal-insulator transition (MIT) in vanadium dioxide ($VO_2$)[1–3] has the potential to lead to a number of disruptive technologies, including ultra-fast data storage, optical switches, and transistors which move beyond the limitations of silicon[4–6]. For applications, $VO_2$ films are deposited on crystalline substrates to prevent cracking observed in bulk $VO_2$ crystals across the thermally driven MIT. Near the MIT, $VO_2$ films exhibit nanoscale coexistence[7–9] between metallic and insulating phases, which opens up further potential applications such as memristors, tunable capacitors,[10–13] and optically engineered devices such as perfect absorbers[14]. It is generally believed that the formation of phase domains must be affected to some extent by random processes which lead to unreliable performance in nanoscale MIT based devices. Here we show that nanoscale randomness is suppressed in the thermally driven MIT in sputtered $VO_2$ films; the observed domain patterns of metallic and insulating phases in the vicinity of the MIT in these films behave in a strikingly reproducible way. This result opens the door for realizing reliable nanoscale $VO_2$ devices.**

The phenomenon of phase coexistence is quite broadly observed across strongly correlated condensed matter systems, occurring for example, in the high-$T_c$ superconducting cuprates[15,16], the colossal magnetoresistive manganites[17–19], as well as the oxides of vanadium[7,20–23]. Highly ordered patterns[15,16,19,21,23] result in response to long range interactions. Generally speaking, the spatial periodicity of these patterns ranges from very small (nanometers or less)[15,16] for strong interactions, such as Coulomb interaction between domains, to hundreds of nanometers or more for somewhat weaker interactions[19,21–23] such as long-range elastic mismatch with a substrate. The increasing periodicity as the interaction strength decreases is attributed to the free energy cost of forming the boundary between domains. In this situation, random processes, such as fluctuations between phases[23] or irreproducibility in the domain pattern upon thermal cycling[22] have been observed. In contrast, amorphous patterns are also commonly observed when imperfections disrupt the long range interaction.[7,18] Generally speaking, there is sparse experimental data on the reproducibility of these patterns.

Due to the stochastic nature of nucleation of a new phase during a first order phase transition, it was expected that this randomness constituted an inherent challenge to creating reproducible phase transition based devices on the scale of the domain size. Until now, the expectation for these systems was that although some regions may preferentially transition due to inhomogeneity and defects, some significant degree of randomness was unavoidable. For example, upon thermal cycling, most but not all of the domain patterns re-emerged in a manganite system exhibiting a first-order phase transition[18]. Strictly speaking, the primary result of our work is that it is *possible* to realize completely reproducible metal-insulator phase domain patterns in a VO$_2$ film. Insight gained through such work could be applied across the entire range of similar correlated materials, whose novel phase transitions have tremendous potential for technological impact.

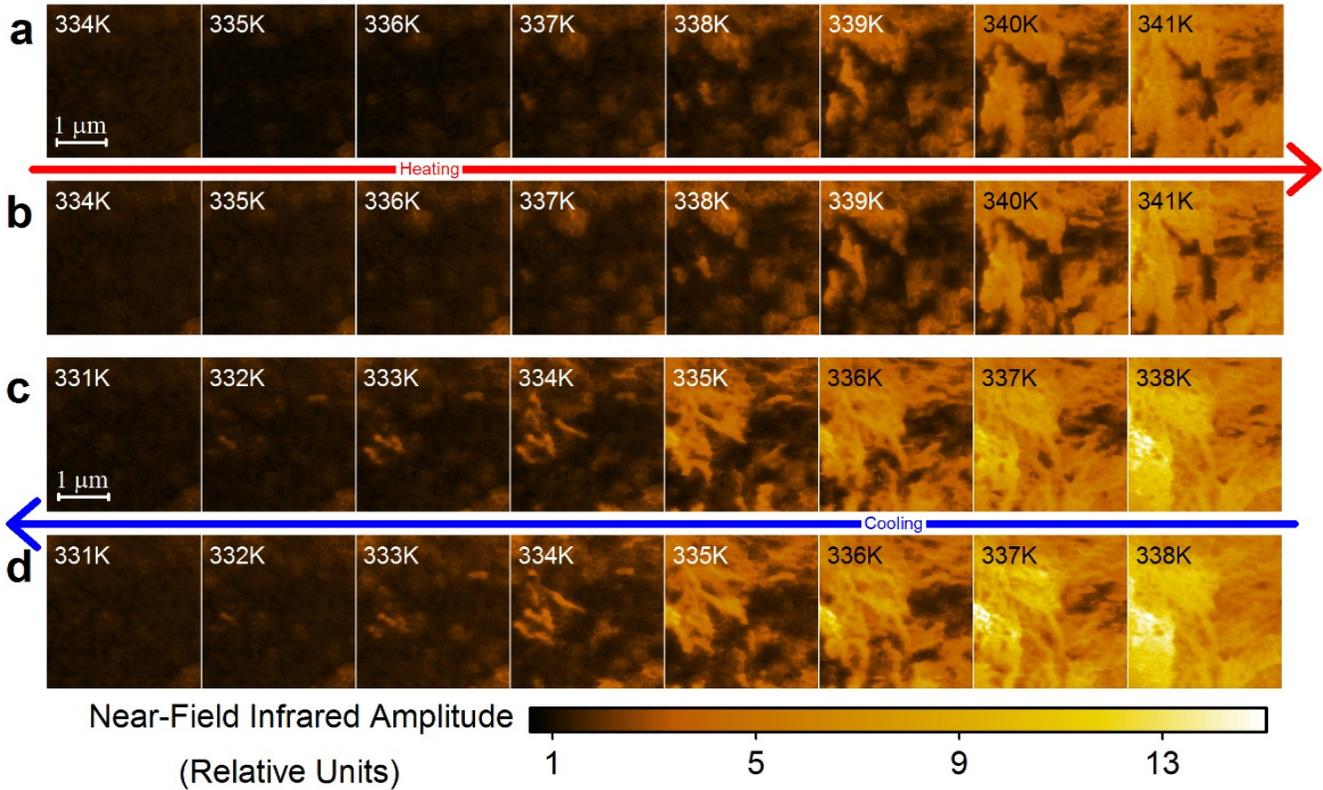

**Figure 1| Reproducible phase coexistence in the VO$_2$ film with temperature.** Near-field infrared amplitude images of the same region obtained with infrared light (λ≈10.6μm). Higher infrared amplitude signals correspond to metallic regions, while lower signals correspond to insulating regions. The signals are normalized to the average signal of the completely insulating 329 K image (not shown). Rows **a** and **b** show separate heating runs. Rows **c** and **d** show separate cooling runs.

In this Letter, we employ scattering-type scanning near-field infrared microscopy (s-SNIM) to image the patterns formed by the coexisting me tallic and insulating domains. We achieve unprecedented spatial stability over a broad temperature range for repeated heating and cooling runs. Representative s-SNIM images obtained in the same spatial region are presented in Fig. 1. Each row represents a separate heating or cooling run through the phase coexistence regime (see Methods for additional information). The nucleation and growth of phase domain patterns are reproducible for the heating runs and the cooling runs respectively. However, somewhat different patterns are observed in the heating runs compared to the cooling runs. As has been seen previously in polycrystalline VO$_2$ films, the MIT is percolation-type in which phase domains first nucleate and then grow in amorphous, fractal-like patterns[7,20,8]. The patterns are static and stable in time, provided the temperature is held constant. While the shapes of these phase domains are reminiscent of those observed in random percolation models of phase transitions, the fact that these domains nucleate and grow in the same way on separate runs through the MIT is evidence that deterministic effects alone dictate the domain patterns. It follows that the domain patterns depend on factors such as grain boundaries, relative orientation of grains, impurities, defects, and dislocations, which are "quenched", or frozen into the film at the time of growth. While the variance introduced by this "quenched disor der" can be treated as random in theoretical models[24], this does not imply that phase domain formation is necessarily a probabilistic (or random) physical process.

To obtain further insight into the domain pattern formation, we report a non-monotonic temperature cycle through the phase coexistence region in a common area of the film in Fig. 2. Interestingly, despite the phase domain patterns on the two heating portions of the cycle being quite different, the same pattern reemerges at 341K. Hence it is not necessary to exit the coexistence regime to reset the memory of the system.

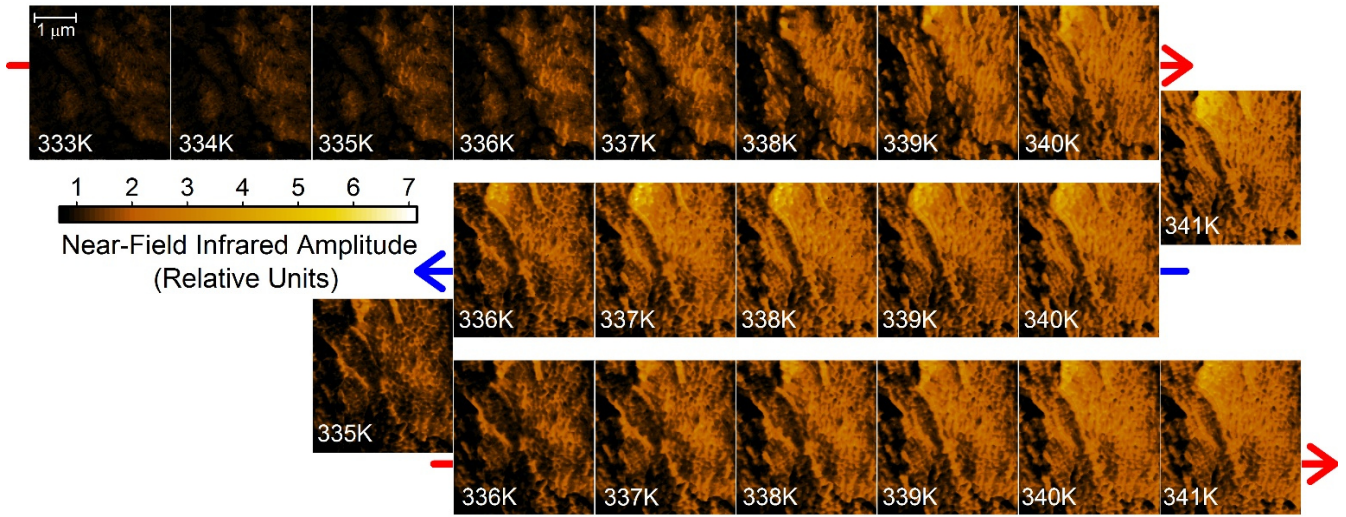

**Figure 2| Dependence on thermal history.** S-SNIM images of a particular region of the VO$_2$ film. The near-field infrared amplitude is normalized to that of the insulating phase in each image. The area scanned here is different from the one shown in Figure 1. Images are obtained during a non-monotonic temperature cycle through the phase coexistence regime. Arrows denote either heating (red) or cooling (blue).

We emphasize that nucleation occurs at precisely the same location in each uni-directional, monotonic temperature excursion across the MIT (See Fig. 3 b,c,d, and e). As the sample temperature crosses the equilibrium temperature, where the free energies of both phases are equal, it is thermodynamically preferred to form a domain ($\Delta G_{Domain}<0$). Kinetically, however, the always positive strain and interfacial free energy terms oppose the formation of domains below a critical size (r*). Thus, any new domain must pass (tunnel) through the nucleation barrier ($\Delta G^*$). This tunneling is an inherently stochastic process; Nucleation occurs at each site in a given time interval with a probability ($P_N$) proportional to $e^{-\Delta G^*_{local}/k_BT}$ (See Fig 3h)[25]. In contrast to homogeneous nucleation, which occurs in the homogeneous bulk, heterogeneous nucleation occurs at some inhomogeneity such as a defect or grain boundary. In heterogeneous nucleation, the size, shape and free energy of forming the critical nucleus can be altered significantly thereby reducing the barrier to nucleation. Nevertheless, as long as a barrier to nucleation exists, the process is inherently stochastic. In the analysis and discussion that follows, we explain that the highly reproducible patterns observed here are due to barrierless nucleation, and deduce qualitative features of the phase transition kinetics.

Quenched disorder locally alters the free energy balance between the phases[26]. As a result, the *local* temperature where the free energy of both phases is equal ($T^{local}_{eq}$) is shifted from the bulk value. As neither phase is thermodynamically preferred at this temperature, some degree of superheating or undercooling is in general necessary to overcome the nucleation barrier. It is natural to consider an elementary (local) hysteresis loop, where each pixel has both a heating and cooling transition temperature (see Methods). Equivalently, each pixel has a hysteresis width ($\Delta T_c$) and an average of the cooling and heating transition temperature ($T^{avg}_c$). The latter is essentially a measure of $T^{local}_{eq}$, if one assumes that the local degree of undercooling and overheating is the same ($\frac{1}{2}\Delta T_c$). In contrast, $\Delta T_c$ is related to the local nucleation barrier. While one could consider a more complicated model, which includes interactions such as strain between domains[27], the re-emergence of the same 341K pattern from different histories (Fig. 2) indicates that the simpler model is sufficient.

We note that nucleation occurs at sites where the local $\Delta T_c$ is suppressed (See Fig 3f), which confirms that the nucleation barrier is greatly reduced, if not completely removed, at these sites. At such a site, the MIT can proceed along a barrierless path, and hence occurs completely deterministically (see Fig. 3h). As one might expect, on heating, nucleation occurs where the local equilibrium temperature ($T_c^{avg}$) is relatively low. Conversely, on cooling, nucleation occurs where $T_c^{avg}$ is relatively high (See Fig. 3 b,c,d,e,f and g).

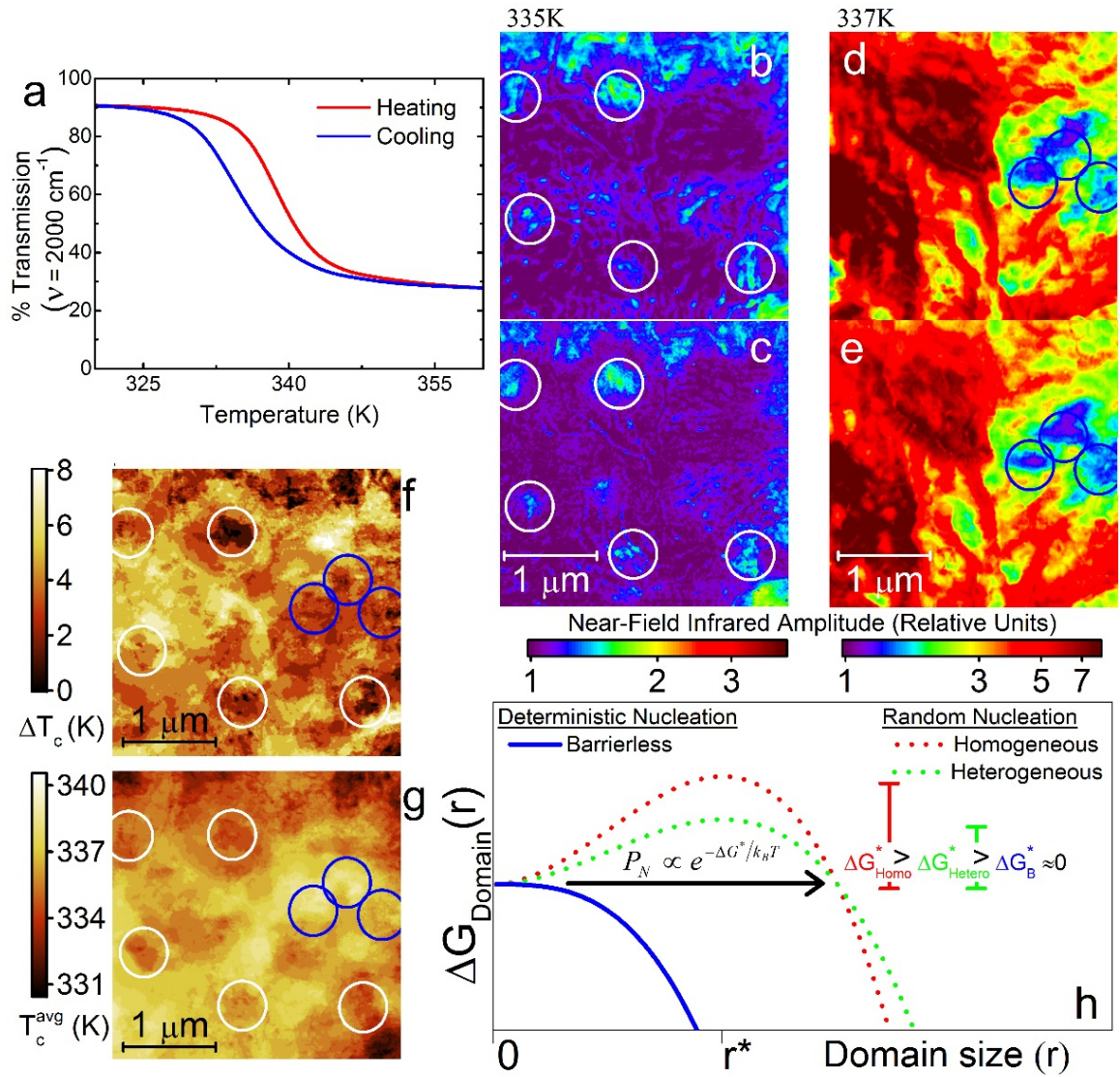

**Figure 3 | Nucleation kinetics of VO2. a,** Macroscopic thermal hysteresis loop measured via infrared transmission through the film-substrate system. **b,c,d,e** s-SNIM images demonstrating the nucleation sites on heating (**b** and **c**), and cooling (**d** and **e**) of the same area as shown in Fig. 1. **g** and **h,** Local hysteresis width ($\Delta T c$) and local phase equilibrium temperature ($T_c^{avg}$) for the area shown in b,c,d,e (See Methods). White (blue) circles in **b,c,d,e,f,g** serve to guide the eye to some of nucleation sites which occur on heating (cooling). **h,** Schematic of the free energy landscape for a domain of characteristic linear dimension *r* for different types of nucleation sites. Note that here we make the distinction between heterogeneous nucleation and "barrierless nucleation". Barrierless nucleation is a special case of heterogeneous nucleation where, unlike the more general case, the nucleation barrier is completely removed, and nucleation thus occurs deterministically.

Both $\Delta T_c$ and $T_c^{avg}$ contribute to the shape of the thermal hysteresis loop observed in macroscopic measurements of the MIT. Inhomogeneity in $T_c^{avg}$ can prevent the propagation of the new phase, resulting in a broader transition as seen in the infrared transmission measurement (Fig. 3a). The variation observed on the microscopic scale is ~9K. That the full breadth of the transition is larger, ~15K, is attributed to the finite field of view of the s-SNIM images. In contrast, macroscopic VO2 single crystals have very sharp transitions, and don't exhibit phase coexistence unless subject to external strain[28–32]. In such crystals, the hysteresis width is set by the smallest local $\Delta T_c$.

Interestingly, several authors have reported that there is a correlation between the size of VO2 single crystals and the width of the hysteresis. It is found that the width of the hysteresis of the MIT can be increased greatly, to as much as 35K for single domain VO2 nanoparticles[33,34]. The hysteresis width is systematically lessened in nano-particles with increasing size and number of grain boundaries[34]. This trend holds for free-standing $VO_2$ crystals, from the somewhat larger $VO_2$ "nanorods"[35] - which have a hysteresis width of approximately 10K - to millimeter scale free-standing $VO_2$ crystals which have hysteresis widths of approximately 2K[36]. A small single domain $VO_2$ nanoparticle is quite likely to not contain a barrierless nucleation site. The huge hysteresis width observed in these crystallites is a clear indication that the nucleation barrier at other sites is so large that these stochastic nucleation processes occur rarely. In contrast, nucleation is *functionally deterministic* in sputtered $VO_2$ films: nucleation occurs reliably - only at the barrierless sites - each time the temperature crosses $T_{eq}^{local}$.

We cannot make definitive statements about what specific local factors create the nucleation sites based on the present work. While there is experimental evidence that grain boundaries play some role[20,34], it is possible that a grain boundary alone is insufficient. Some other factor, perhaps below the length scale of this measurement, is necessary to create a nucleation site. One such factor which has been suggested previously via DFT+U calculations, is an oxygen vacancy at a grain boundary[34]. Perhaps a single oxygen vacancy is insufficient and a critical concentration of oxygen vacancies per unit volume is required for nucleation. There is reason to believe that the multi-grain structure of the present sample is essential to deterministic behavior, as some evidence of randomness has been seen previously in $VO_2$ samples where this is not the case: Dramatic spontaneous symmetry breaking has been attributed to the structural phase transition accompanying the MIT in a particularly clean $VO_2$ film[22]. Furthermore, differences in the metal-insulator domain pattern was reported on cooling $VO_2$ nanorods, but only at one of many reported temperatures[37].

To conclude, we have shown that the phase domain patterns which form during the thermally driven MIT in a $VO_2$ film can behave in a completely deterministic way. It is clear from the present result that quenched disorder can be used to reliably control the spatial distribution of phase domains. Interestingly, nanoscale inhomogeneity in the ultra-fast optically driven MIT suggests that this conclusion can be generalized to the dc electric field driven MIT as well[37,38]. Our work lays the foundation for further exploration into reliable nanoscale $VO_2$ electronic and photonic devices.

## Methods

Scattering type scanning near field infrared microscopy (s-SNIM) images were taken over a number of runs through the metal-insulator transition. To investigate the reproducibility of the patterns produced by the phase domains, raster scans were taken in the same area on the film in repeated thermal runs. We performed two heating runs in three different regions of the film for a total of six heating runs. In none of the regions was any evidence of randomness observed. Three cooling runs were performed at the location shown in Fig. 1 to compare the patterns to the heating runs and verify the repeatability in the cooling direction. In all of these runs, the sample was first brought to a temperature completely outside of the phase coexistence regime where it was fully metallic (for cooling) or fully insulating (for heating). The non-monotonic run presented in Fig. 2 was taken in yet another area.

In s-SNIM, infrared light is scattered from a metal coated tip of a tapping mode atomic force microscope (AFM)[39]. The scattered light contains information about the near-field interaction between the tip and the region of the sample immediately below the tip. This procedure allows for the simultaneous collection of topographic and optical images of a given region of the film, with resolution on the order of the AFM tip radius, approximately 15 to 20 nm. The pseudo-heterodyne detection scheme and demodulation of the optical signals at the third harmonic are used to reduce various background contributions[39,40]. In s-SNIM, the local topography can influence the signal, i.e. the signal is generally higher in a valley than on a topographic peak. This surface roughness (~3nm RMS) induced s-SNIM contrast can be seen in the low temperature (purely insulating) images, and causes a variation of around 10%. The wavelength of 10.6 μm used in this work is within the band gap of insulating $VO_2$, and above the infrared active phonon region. The large change in optical constants across the MIT at this wavelength results in significant infrared contrast between metallic regions (high infrared signal) and insulating regions (low infrared signal), much greater than that caused by the topography. That the s-SNIM images presented here are due to the MIT is clear because they are temperature dependent while the topography is temperature independent.

To determine the local transition temperatures used in Figure 3, we employ a threshold of 1.45 for the normalized infrared amplitude, above which the pixel is considered to be metallic. We define the local transition temperature as the temperature where the signal first crosses the threshold for both the heating and cooling directions. The observed local characteristics (See Fig. 3 f and g) – and thus our conclusions – do not change appreciably for a broad range of reasonable thresholds. Appropriate thresholds are those that exceed the topography induced contrast, but are still low enough to capture the subtle contrast near nucleation.

In this work, we study the thermally driven MIT in a 45nm $VO_2$ film grown by RF sputtering on (001) sapphire. In such films, the monoclinic angle (~123°) of the $M_1$ phase tends to align in plane along the 120° angles of the hexagonal sapphire lattice. Thus, grains in this film will prefer one of six possible orientations, differing by an out of plane rotation, due to the rotational symmetry of the hexagonal sapphire substrate[41]. Sputtered films tend to have some additional compressive in-plane strain due to an effect known as "shot peening[42]." Compressive strain along the $a_{M1}$ ($c_r$) axis stabilizes the rutile phase, resulting in a somewhat lower $T_c$. These films are ideal for this work in that they have distinct topographic features, in the form of "valleys", which can be used to ensure that images are consistently taken in the same area. Thermal drift was minimized by placing the sample and measurement region at the center of a circular heating stage, which was designed in-house. A silicon diode based thermometer and a resistive heating element were mounted on the stage and these components along with a Lakeshore model 335 temperature controller were employed for thermal management. A slow temperature ramp rate ~0.2K/minute was used to minimize overshooting (<0.1K) of the set-point temperature. Minute differences in this overshoot can result in an image that appears somewhat more progressed in one run as opposed to another. Once the set point temperature is obtained, it is held stable within 0.1 K for the full duration of the scan.

## Acknowledgements

MMQ acknowledges support from NSF DMR (grant # 1255156).

## Author Contributions

TJH, DJL, and SLW performed the infrared experiments. MMQ conceived the project and supervised the infrared experiments. The $VO_2$ films were synthesized by TS and BJK, and HTK supervised this part of the project. TJH, TS, HTK, and MMQ wrote the manuscript.